# Presentation of tracing sensation with different roughness by using disk with uneven surface

Soma Kato, Izumi Mizoguchi, and Hiroyuki Kajimoto

*The University of Electro-Communications, Chofu, Japan*

(Email: kato@kaji-lab.jp)

**Abstract ---** Tracing sensation plays an important role in discriminating friction and texture of objects. We have been studying a method of presenting the tracing sensation by contacting the center of a rotating disk with the fingertip. Although this method can reduce the size and complexity of the presentation device compared to conventional methods, it can only present single type of roughness. In this study, we examined a method to dynamically modulate the roughness by devising the surface shape of the disk. We designed two types of disks: slit disks and milling disks. Experiments revealed that the perceived roughness of these disks can be modulated by altering the rotation speed and direction.

Keywords: haptic display, rotating disk, tracing sensation

## 1 Introduction

We perform tracing motions on a daily basis to perceive the texture and friction of objects. The sensation that occurs during this process is referred to as the tracing sensation. The presentation of the tracing sensation plays an important role in enhancing the quality of the virtual reality experience.

Conventional methods for presenting the tracing sensation are generally based on reproducing the skin deformation during tracing. These methods are mainly classified into those that change the coefficient of friction [1][2][3][4], those that move the contact surface in the shear direction [5][6][7], and those that use the rotation of rollers, disks, or belts [8][9][10]. Among these methods, the last one is considered to provide a highly realistic tracing sensation because it uses actual objects. On the other hand, these methods tend to require large or complicated equipment.

In our previous study, we have proposed a method of presenting a tracing sensation by contacting the center of a rotating disk with the fingertip [11][12] (Fig.1 ). Compared to conventional methods, this approach has the advantage of a much smaller size. On the other hand, it has the limitation of fixed texture being presented, which restricts its potential applications.

In this study, we extend our previous approach and propose a method to dynamically modulate the perceived roughness during tracing motion by modifying the surface shape of the rotating disk. By applying bumps to the surface of the disk, it is considered that the skin deformation pattern of the fingertip can be modulated by changing parameters such as rotation speed and direction, which may alter the perceived roughness.

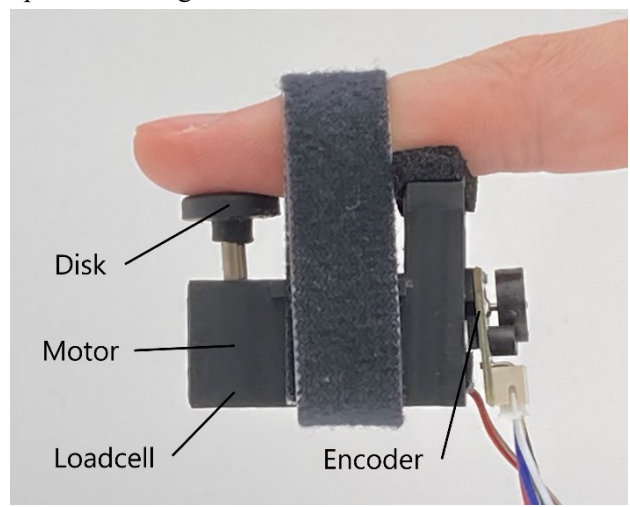

Fig.1 Overview of the device for presenting the tracing sensation

On the other hand, a previous study suggested that texture perception does not depend on tracing speed [13]. Our previous studies also supported this finding, i.e. the rotation speed of the disk did not affect perceived roughness. On the other hand, these observations were mainly for fine textures. Since this study focuses on rough textures or bumps where tracing velocity on the skin surface is perceptible, we considered that the perceived roughness can be affected by rotating speed.

## 2 IMPLEMENTATION

### 2.1 Device

The configuration of the device is the same as in our previous study [11][12] (Fig.1 ) and mainly consists of a DC motor to rotate the disk, an encoder to control the disk, and a load cell (HSFPAR003A, Alps Alpine) to measure the pressing force on the disk. By using a motion capture system, the rotation speed of the disk is controlled according to the tracing speed.

### 2.2 Disks

The two types of disks we have designed are shown in Fig. 2 and Fig.3 .

The slit disk (Fig.2 ) is considered to generate vibrations proportional to the number of slits and the rotational speed of the disk. This suggests that a larger number of slits and a higher rotational speed may result in a smoother sensation.

The milling disk (Fig.3 ) is inspired by a milling tool, which has sharp edges that removes workpiece. Due to its structure, it is expected to be perceived as rough during clockwise rotation and smooth during counterclockwise rotation. To investigate the effect of the angle of the step, three types of disks with different inclination ratios were designed.

## 3 EXPERIMENT

In order to verify the effectiveness of the two types of disks, we conducted two experiments. Experiment 1 investigated the effect of rotation speed on perceived roughness using the slit disks. Experiment 2 investigated the effect of rotation direction on perceived roughness in the same way using the milling disks.

### 3.1 Setup

The experimental setup was identical for Experiments 1 and 2.

A linear slider (RSA0N11M9A0K, Alps Alpine) was used to control the tracing speed and distance during stimuli presentation. In the experiment, participants traced three round trips from one end of the linear slider to the other in response to an audio cue that sounded every second. Since the actual measured distance traveled by the linear slider was 95 mm, the tracing speed was 95 mm/s (assuming constant velocity motion).

Sandpaper was used as a measure of the sensation presented by the device. A total of seven types of sandpaper (#60, #120, #240, #400, #800, #1500, and #3000) were used in the experiment. In the experiment, the sandpapers were assigned numbers from 1 to 7 in order of smooth to rough. These sandpapers were replaced with new ones for each participant to minimize the effect of wear.

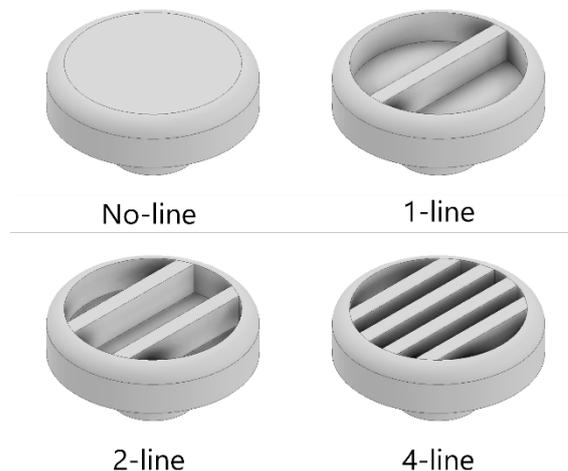

Fig.2  4 types of slit disks.

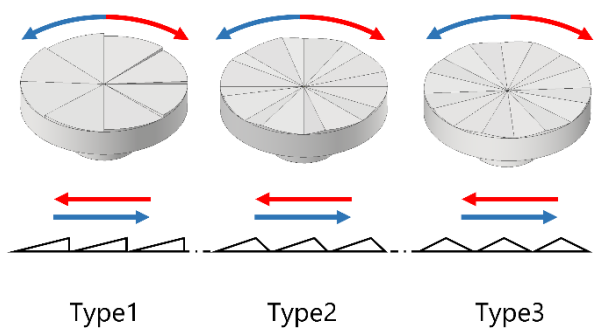

Fig.3  3 types of milling disks

Initially, participants wore the device on their index finger and performed tracing movements in response to the audio cues while placing their finger on the linear slider. During this time, the disk in contact with the fingertip rotated, presenting the tracing sensation. Participants then removed the device and selected a piece of sandpaper that most closely resembled the sensation presented by the device. This procedure was repeated for each condition.

Seven participants (6 males, 1 female, ages 21-24) participated in Experiment 1 and five participants (5 males, ages 21-27) participated in Experiment 2.

### 3.2 Experimental Conditions of Experiment 1

In Experiment 1, we investigated the effect of rotation speed on perceived roughness for four types of disks (1-line, 2-line, 4-line, and No-line): three types of slit disks with different numbers of lines, and one disk without a slit. The rotation speed is the same as in our previous study [11], and three conditions were used: the circumferential speed of the disk is equal to 36% of the tracing speed (Middle condition), and 0.5 times (Low condition) and 2 times (High condition) of the Middle condition.

### 3.3 Results of Experiment 1

The results of Experiment 1 are shown in Fig.4 . The aligned rank transform analysis of variance (ART-ANOVA) [14] was used for the tests. The results for the 1-line condition and the 2-line condition indicate that the perceived roughness increased as the rotational speed increased, a result contrary to our hypothesis. A possible explanation for this is that the vibration simply became more intense as the speed increased, which could have been interpreted as rougher. On the other hand, this tendency weakened as the number of lines increased, suggesting that the relationship between rotational speed and perceived roughness could be altered by increasing the number of lines or by using a different shape.

### 3.4 Experimental Conditions of Experiment 2

In Experiment 2, the effects of rotation direction on perceived roughness were investigated for three types of milling disks (Type 1, Type 2, and Type 3). The direction of rotation was either clockwise (Clockwise condition) or counterclockwise (CounterClockwise condition), for a total of two conditions.

### 3.5 Results of Experiment 2

The results of Experiment 2 are shown in Fig.5 . Statistical tests were not conducted due to the small number of subjects. The results indicate that the Type 1 disks were perceived as rougher in the Clockwise condition than in the CounterClockwise condition. However, the difference was slight, which may be due to the lack of depth of the bump. It may be possible to change the roughness perception more significantly by optimizing the depth in the future.

### 4 CONCLUSION

In this study, we designed two types of disks, slit disks and milling disks, and proposed a method for modulating perceived roughness while presenting a tracing sensation by using these disks. In addition, we conducted an experiment to verify the relationship between the sensation of the stimuli presented by the device and the perceived roughness, using sandpaper of different roughness as an evaluation index. The results showed that the perceived roughness modulates by altering the rotation speed or direction.

In out next step, we will continue to optimize surface shapes and experimental devices based on the findings of this research, while also investigating different approaches to roughness modulation, such as electrostatic tactile sensation.

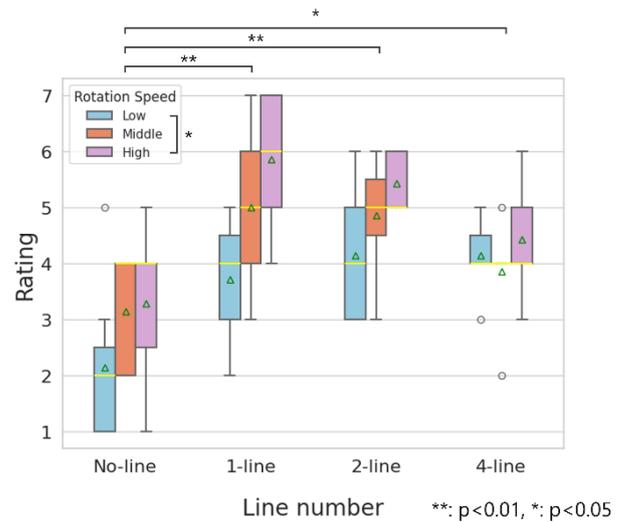

Fig.4 The results of Experiment 1. The yellow line indicates the median, and the green triangle indicates the mean.

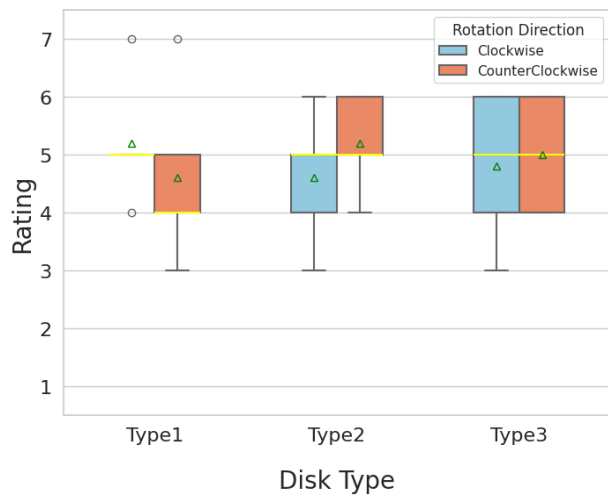

Fig.5 The results of Experiment 2. The yellow line indicates the median, and the green triangle indicates the mean.

ACKNOWLEDGEMENT

This research was supported by Support Center for Advanced Telecommunications Technology Research (SCAT) and JSPS KAKENHI Grant Number JP20H05957

REFERENCES

[1] O. Bau, I. Poupyrev, A. Israr, and C. Harrison, "TeslaTouch: electrovibration for touch surfaces," in Proceedings of the 23nd annual ACM symposium on User interface software and technology, NY, USA, 2010

[2] C. Basdogan, F. Giraud, V. Levesque, and S. Choi, "A Review of Surface Haptics: Enabling Tactile Effects on Touch Surfaces," IEEE Transactions on Haptics, vol. 13, no. 3, pp. 450–470, 2020

[3] Y. Vardar, B. Güçlü, and C. Basdogan, "Effect of Waveform on Tactile Perception by Electrovibration Displayed on Touch Screens," IEEE Transactions on Haptics, vol. 10, no. 4, pp. 488–499, 2017

[4] E. AliAbbasi, Mr. A. Sormoli, and C. Basdogan, "Frequency-Dependent Behavior of Electrostatic Forces Between Human Finger and Touch Screen Under Electroadhesion," IEEE Transactions on Haptics, vol. 15, no. 2, pp. 416–428, 2022

[5] V. Yem, R. Okazaki, and H. Kajimoto, "FinGAR: combination of electrical and mechanical stimulation for high-fidelity tactile presentation," in ACM SIGGRAPH 2016 Emerging Technologies, NY, USA, 2016

[6] C. Ho, J. Kim, S. Patil, and K. Goldberg, "The Slip-Pad: A haptic display using interleaved belts to simulate lateral and rotational slip," in 2015 IEEE World Haptics Conference (WHC), 2015.

[7] S. B. Schorr and A. M. Okamura, "Three-Dimensional Skin Deformation as Force Substitution: Wearable Device Design and Performance During Haptic Exploration of Virtual Environments," IEEE Transactions on Haptics, vol. 10, no. 3, pp. 418–430, 2017

[8] E. Whitmire, H. Benko, C. Holz, E. Ofek, and M. Sinclair, "Haptic Revolver: Touch, Shear, Texture, and Shape Rendering on a Reconfigurable Virtual Reality Controller," in Proceedings of the 2018 CHI Conference on Human Factors in Computing Systems, NY, USA, 2018

[9] M. J. Kim, N. Ryu, W. Chang, M. Pahud, M. Sinclair, and A. Bianchi, "SpinOcchio: Understanding Haptic-Visual Congruency of Skin-Slip in VR with a Dynamic Grip Controller," in Proceedings of the 2022 CHI Conference on Human Factors in Computing Systems, NY, USA, 2022

[10] P. Zhang, M. Kamezaki, Y. Hattori, and S. Sugano, "A Wearable Fingertip Cutaneous Haptic Device with Continuous Omnidirectional Motion Feedback," in 2022 International Conference on Robotics and Automation (ICRA), 2022

[11] S. Kato, Y. Suga, I. Mizoguchi, and H. Kajimoto, "Presentation of a Tracing Sensation by Means of Rotation Stimuli," in 2024 IEEE Haptics Symposium (HAPTICS), 2024

[12] S. Kato, Y. Suga, I. Mizoguchi, and H. Kajimoto, "Presentation of Tracing Sensation through Combination of Disk Rotation and Vibration" in Eurohaptics conference 2024, 2024

[13] Z. M. Boundy-Singer, H. P. Saal, and S. J. Bensmaia, "Speed invariance of tactile texture perception," J Neurophysiol, vol. 118, no. 4, pp. 2371–2377, 2017

[14] J. O. Wobbrock, L. Findlater, D. Gergle, and J. J. Higgins, "The aligned rank transform for nonparametric factorial analyses using only anova procedures," in Proceedings of the SIGCHI Conference on Human Factors in Computing Systems, NY, USA, 2011